\documentclass[a4paper]{jpconf}

\newcommand{\mean}[1]{\left<#1\right>}
\newcommand{\means}[1]{\langle#1\rangle}

\usepackage{bm}
\usepackage{amsmath,amssymb}
\usepackage[usenames]{color}

\usepackage{graphicx}

\def\url#1{}
\usepackage{cite}

\begin{document}
\title{Spin correlation and Majorana spectrum in chiral spin liquids in a decorated-honeycomb Kitaev model}

\author{Joji Nasu$^{1}$ and Yukitoshi Motome$^{2}$}

\address{$^{1}$Department of Physics, Tokyo Institute of Technology, Ookayama, 2-12-1, Meguro, Tokyo 152-8551, Japan}
\address{$^{2}$Department of Applied Physics, University of Tokyo, Hongo, 7-3-1, Bunkyo, Tokyo 113-8656, Japan}

\ead{nasu@phys.titech.ac.jp}

\begin{abstract}
Temperature evolution of the spin correlation and excitation spectrum is investigated for a Kitaev model defined on a decorated honeycomb lattice by using the quantum Monte Carlo simulation in the Majorana fermion representation.
The ground state of this quantum spin model is given by two kinds of chiral spin liquids: one is topologically trivial with Abelian anyon excitations, and the other is topologically nontrivial with non-Abelian anyon excitations.
While lowering temperature, the model exhibits several crossovers in the paramagnetic state, which originate from the fractionalization of quantum spins into Majorana fermions, in addition to a phase transition associated with time reversal symmetry breaking.
 We show that the spin correlation develops around the crossover temperatures, whereas it shows a slight change at the critical temperature, as in other Kitaev-type models.
We also calculate the excitation spectrum in terms of Majorana fermions, and find that the excitation gap in the non-Abelian phase is fragile against thermal fluctuations of the $Z_2$ fluxes, while that in the Abelian phase remains open.
\end{abstract}

\section{Introduction}

%% Kitaev model

The quantum spin models that can be solved exactly have often provided a clear insight into quantum many-body physics~\cite{Bethe31,Huthen38,Baxter82}. 
For instance, the Affleck-Kennedy-Lieb-Tasaki model~\cite{PhysRevLett.59.799} and the Shastry-Sutherland model\cite{SRIRAMSHASTRY19811069} have been intensively studied for establishing the valence-bond solid picture in one and two dimensions, respectively.
The Kitaev model is one of such solvable models, which is composed of $S=1/2$ spins with bond-dependent interactions on a two-dimensional honeycomb lattice~\cite{Kitaev2006}.
A striking feature in the Kitaev model is that it possesses a macroscopic number of conserved quantities, 
which allow us to map the model onto a free Majorana fermion system coupled to $Z_2$ variables. 
The ground state is exactly shown to be a quantum spin liquid, which is one of the most intriguing subjects in strongly correlated electron systems~\cite{ISI:000275366100033}.
The exact solvability is inherited in other lattices if they are tri-coordinated: 
indeed, many generalizations of the Kitaev model have been studied extensively in both two and three dimensions~\cite{PhysRevLett.99.247203,PhysRevB.79.024426,PhysRevB.76.180404,PhysRevB.89.235102,PhysRevB.90.205126,PhysRevB.89.045117,PhysRevLett.114.116803,PhysRevB.89.115125,PhysRevLett.113.197205}.

Among the tri-coordinate lattices, the Kitaev model on a decorated honeycomb lattice has been of great interest as the ground state is a chiral spin liquid (CSL), where the time reversal symmetry is broken without magnetic orders~\cite{Kitaev2006,PhysRevLett.99.247203,PhysRevB.81.060403}.
This lattice is obtained from the honeycomb lattice by replacing each lattice site by a triangle [see Fig.~\ref{lattice}(a)].
The model has two types of conserved quantities: one is on each triangle and the other is on each dodecagon.
The former corresponds to the chirality degree of freedom, and hence, it plays a key role in stabilizing the CSL ground state with time reversal symmetry breaking.
Interestingly, the model exhibits two types of CSLs while changing the exchange parameters:
 the topologically-trivial CSL with Abelian anyon excitations and the topologically-nontrivial CSL with non-Abelian anyon excitations~\cite{PhysRevLett.99.247203}.
While the exact solution is limited for zero temperature ($T$), the thermodynamic properties were also investigated numerically~\cite{PhysRevLett.115.087203,Nasu2015pre}.
There is a finite-$T$ phase transition between the CSLs and paramagnet, and the nature of the phase transition is dependent on the topological nature of the CSLs.
Nevertheless, thermodynamic properties of CSLs remain elusive. In particular, the magnetic properties and the dynamics at finite $T$ have not been studied in detail thus far.
These are important not only for illuminating the nature of exotic CSL states but also examining the stability of anyon excitations that are useful in quantum computation.

%% In this paper

In this paper, we investigate the spin correlation and excitation spectrum at finite $T$ in the Kitaev model on the decorated honeycomb lattice by performing the quantum Monte Carlo (QMC) simulation in the Majorana fermion representation~\cite{PhysRevLett.113.197205}.
We find that the nearest-neighbor (NN) spin correlation develops around high-$T$ crossovers yielded by the entropy release of the itinerant Majorana fermions.
Moreover, computing the density of states (DOS) of the Majorana fermions, we show a close relationship between the crossover temperatures and the $T$ evolution of the Majorana spectrum. 
We also reveal that the low-energy gap in the topologically-nontrivial CSL is smeared out above the critical temperature by thermal fluctuations of $Z_2$ fluxes associated with the localized Majorana fermions.
In contrast, the gap in the topologically-trivial CSL remains robust against the thermal fluctuations.

\section{Model}

\begin{figure}[t]
\begin{center}
\includegraphics[width=0.5\columnwidth,clip]{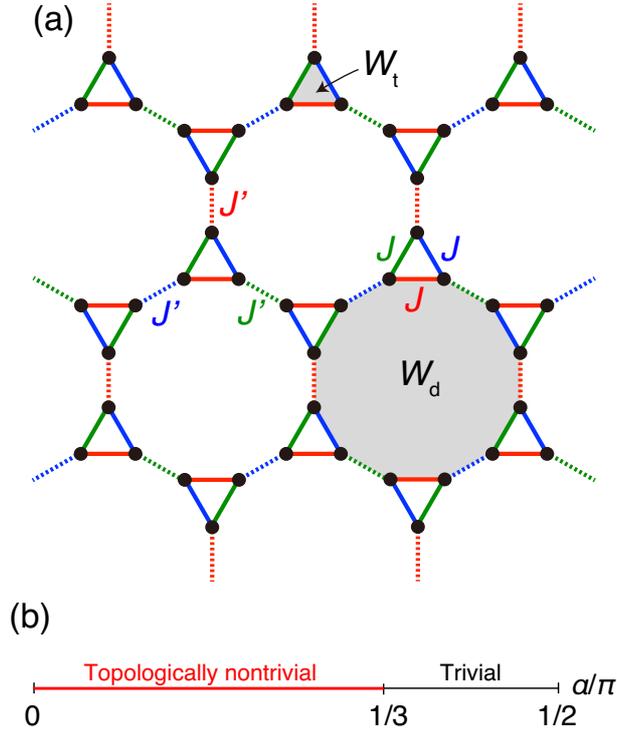}
\caption{
 (a) Schematic picture of the Kitaev model in Eq.~(\ref{eq:1}) on a decorated honeycomb lattice. The solid (dotted) blue, green, and red bonds represent the intra(inter)-triangle $x$, $y$, and $z$ bonds with the exchange constant $J$ ($J'$), respectively.
 $W_{\rm t}$ and $W_{\rm d}$ represent the conserved quantities (fluxes) defined on each triangle and dodecagon, respectively.
 (b) Ground state phase diagram as a function of $\alpha$; we take $J=\cos\alpha$ and $J'=\sin\alpha$.
 There are two CSL phases with topologically trivial and nontrivial Majorana fermion bands, which meet at the critical point $\alpha_c/\pi=1/3$.
 }
\label{lattice}
\end{center}
\end{figure}

The Kitaev model on the decorated honeycomb lattice is defined by the Hamiltonian~\cite{Kitaev2006,PhysRevLett.99.247203}:
\begin{align}
 {\cal H}=-J\sum_{\gamma=x,y,z}\sum_{\means{jk}_\gamma}
 \sigma_j^\gamma\sigma_k^\gamma
 -J'\sum_{\gamma'=x,y,z}\sum_{\means{jk}'_{\gamma'}}
 \sigma_j^{\gamma'}\sigma_k^{\gamma'},\label{eq:1}
\end{align}
where $\sigma_j^\gamma$ is the $\gamma(=x,y,z)$ component of the Pauli matrices representing an $S=1/2$ spin localized on site $j$;
$\means{jk}_\gamma$ represents a NN $\gamma$ bond within a triangle with the exchange constant $J$ and $\means{jk}'_{\gamma'}$ represents a NN $\gamma'$ bond connecting triangles with the exchange constant $J'$ [see Fig.~\ref{lattice}(a)].
We parameterize the exchange constants by introducing the parameter $\alpha$ as $J=\cos\alpha$ and $J'=\sin\alpha$ ($0 < \alpha < \pi/2$, i.e., $J>0$ and $J'>0$).

By mapping the quantum spins $\sigma_j^\gamma$ onto the fermions by the Jordan-Wigner transformation and representing the fermions by two types of Majorana fermions $c_j$ and $\bar{c}_j$~\cite{PhysRevB.76.193101,PhysRevLett.98.087204,1751-8121-41-7-075001}, the Kitaev model in Eq.~(\ref{eq:1}) is rewritten into the form of
\begin{align}
{\cal H}=iJ \sum_{(jk)_x}c_j c_k
-iJ\sum_{(jk)_y}c_j c_k
-iJ\sum_{(jk)_z}\eta_r c_j c_k
+iJ' \sum_{(jk)'_x}c_j c_k
-iJ'\sum_{(jk)'_y}c_j c_k
-iJ'\sum_{(jk)'_z}\eta_{r} c_j c_k, 
\label{eq:2}
\end{align}
where $\eta_r = i\bar{c}_j \bar{c}_k$ ($r$ is the label of the bond).
Here, $\{\eta_r\}$ are the $Z_2$ conserved quantities taking $\pm 1$, as $\eta_r^2=1$ and they commute with each other and the Hamiltonian.
The summations with $(jk)_\gamma$ and $(jk)'_{\gamma'}$ are taken for the NN pairs on the intra-triangle $\gamma$ and inter-triangle $\gamma'$ bonds with $j<k$, respectively.
Thus, the Kitaev model is described by a quadratic Hamiltonian in terms of Majorana fermions, in which the itinerant Majorana fermions $c$ are coupled with the $Z_2$ variables $\eta$ composed of the localized Majorana fermions $\bar{c}$.
This model has two types of conserved quantities, $W_{\rm t}$ and $W_{\rm d}$, defined on triangular and dodecagon plaquettes, respectively, both of which are termed fluxes. 
They are defined by the product of the Pauli matrices on each plaquette as $W_{\rm t}=\prod_{j=1}^3\sigma_j^{\gamma_j}$ and $W_{\rm d}=\prod_{j=1}^{12}\sigma_j^{\gamma_j}$, where $\gamma_j$ represents the label of the bond that is not included in the plaquette among the three bonds connected to the site $j$.
Equivalently, the fluxes are also written by the product of $\eta_r$ included in the corresponding plaquette.

The ground state of the decorated-honeycomb Kitaev model is given by all $W_{\rm t} = W_{\rm d} = +1$ (zero flux state), namely, all $\eta_r$ being $+1$.
This is a free Majorana fermion system on a uniform background of the $Z_2$ variables.
The spin correlation $\means{\sigma_j^\gamma \sigma_k^\gamma}$ is nonzero only for the NN bonds:
the ground state is a QSL with extremely short-range spin correlations~\cite{PhysRevLett.98.247201}.
As the conserved quantity $W_{\rm t}$ changes its sign by the time reversal operation, the ground state with all $W_{\rm t} = +1$ is a CSL with time reversal symmetry breaking~\cite{Kitaev2006}. 
Interestingly, the current model exhibits two kinds of CSLs: a topologically-trivial CSL with Abelian anyon excitations and a topologically-nontrivial CSL with non-Abelian anyon excitations~\cite{PhysRevLett.99.247203}.
The ground state phase diagram as a function of $\alpha$ is presented in Fig.~\ref{lattice}(b).
There is a critical point at $\alpha_c=\pi/3$, which separates the topologically-trivial CSL for $\alpha > \alpha_c$ and the topologically-nontrivial CSL for $\alpha < \alpha_c$. 
Both CSLs have a nonzero gap in the excitation spectrum, while it goes to zero continuously approaching the critical point from both sides, as shown in Fig.~\ref{gap}.
In the non-Abelian CSL region for $\alpha<\alpha_c$, the system exhibits a chiral edge mode in the gap reflecting the topological nature~\cite{PhysRevLett.99.247203}.

\section{Method}

In order to investigate finite-$T$ properties in the decorated-honeycomb Kitaev model, we perform the QMC simulation in the Majorana fermion representation~\cite{PhysRevLett.113.197205}.
As described in Eq.~(\ref{eq:2}), the Kitaev model is regarded as a noninteracting Majorana fermion system coupled to the classical $Z_2$ variables $\eta_r$ defined on the $z$ bonds.
The partition function is given by
\begin{align}
 Z={\rm Tr}_{\{c_j\}}{\rm Tr}_{\{\eta_r\}}\exp\left[-\beta {\cal H}\right]=\sum_{\{\eta_r=\pm 1\}}\exp\left[-\beta F_f(\{\eta_r\})\right],
\end{align}
where $F_f(\{\eta_r\})=-\beta^{-1}\ln {\rm Tr}_{\{c_j\}}\exp\left[-\beta {\cal H}\right]$ is the free energy of the free Majorana fermion system for a given configuration of $\{\eta_r\}$ and $\beta=1/T$ is the inverse temperature (we set the Boltzmann constant to unity).
In the Monte Carlo (MC) simulation, sequence of configurations of $\{\eta_r\}$ is generated by the Markov chain so as to reproduce the distribution of $\exp\left[-\beta F_f(\{\eta_r\})\right]$.
The free energy $F_f$ is computed by using the exact diagonalization method as shown below [see Eq.~(\ref{eq:F_f})].
The calculations are done for the $6L^2$-site clusters up to $L=10$ with a twisted boundary condition so that the $x$ and $y$ bonds comprise a single one-dimensional chain.
We adopted the replica exchange technique~\cite{doi:10.1143/JPSJ.65.1604} with 16 replicas and performed 40000 MC steps for measurement after 10000 MC steps for thermalization in each replica.

For a given configuration of the $Z_2$ variables $\{\eta_r\}$, the Hamiltonian in Eq.~(\ref{eq:2}) is diagonalized as
\begin{align}
 {\cal H}(\{\eta_r\})=\sum_{\lambda}E_\lambda(\{\eta_r\})\left(f_\lambda^\dagger f_\lambda-\frac{1}{2}\right),
\end{align}
where $E_\lambda$ ($\geq 0$) is the one-particle energy of a fermion described by the creation (annihilation) operator $f_\lambda^\dagger$ ($f_\lambda$).
The free energy of the fermion system is calculated as
\begin{align}
\label{eq:F_f}
 F_f(\{\eta_r\})=\beta^{-1}\sum_{\lambda}\ln\left[2\cosh\frac{\beta E_\lambda(\{\eta_r\})}{2}\right].
\end{align}
The specific heat per site are evaluated by
\begin{align}
 C_v=\frac{\beta^2}{6L^2}\left(\means{E_f^2}-\means{E_f}^2-\mean{\frac{\partial E_f}{\partial \beta}}\right),\label{eq:3}
\end{align}
where $E_f$ is the energy of the fermion system given by
\begin{align}
 E_f(\{\eta_r\})=-\sum_{\lambda}\frac{E_\lambda(\{\eta_r\})}{2}\tanh\frac{\beta E_\lambda(\{\eta_r\})}{2}.
\end{align}
The entropy per site is calculated from the the specific heat as
\begin{align}
 S=\ln 2-\int_{T}^{T_m} \frac{C_v}{T'}dT',\label{eq:7}
\end{align}
where we choose $T_m=10$~($\gg J,J'$).
We also introduce the DOS of the Majorana fermion system at finite $T$, which is defined by
\begin{align}
 D(\omega)=\frac{1}{3L^2}\sum_{\lambda}\means{\delta(\omega-E_\lambda(\{\eta_r\}))}.
 \label{eq:DOS}
\end{align}
Here, we take into account the effect of temperature only through thermal fluctuations of the $Z_2$ variables but not in the Fermi distribution function.
Note that the first two terms in Eq.~(\ref{eq:3}) are the contribution from the $Z_2$ variables and the third term comes from the itinerant Majorana fermions.
The latter contribution is also written by using the DOS as
\begin{align}
 \tilde{C}_v=-\frac{\beta^2}{6L^2}\mean{\frac{\partial E_f}{\partial \beta}}=\frac{1}{2}\int_0^\infty D(\omega) \left[\frac{\beta\omega}{2\cosh(\beta\omega/2)}\right]^2 d\omega.\label{eq:6}
\end{align}
We also calculate the spin correlations for the intra- and inter-triangle NN bonds, which are given by
\begin{align}
 S_{\rm NN}^{\rm intra}=\frac{1}{6L^2}\sum_{\gamma=x,y,z}\sum_{\means{ij}_\gamma}\means{\sigma_i^\gamma\sigma_j^\gamma}=\frac{1}{6L^2}\left(-i\sum_{(jk)_x}\means{c_j c_k}
+i\sum_{(jk)_y}\means{c_j c_k}
 +i\sum_{(jk)_z}\means{\eta_r c_j c_k}\right)\label{eq:4}
\end{align}
and
\begin{align}
  S_{\rm NN}^{\rm inter}=\frac{1}{3L^2}\sum_{\gamma'=x,y,z}\sum_{\means{ij}'_{\gamma'}}\means{\sigma_i^{\gamma'}\sigma_j^{\gamma'}}=\frac{1}{3L^2}\left(-i\sum_{(jk)'_x}\means{c_j c_k}
+i\sum_{(jk)'_y}\means{c_j c_k}
 +i\sum_{(jk)'_z} \means{\eta_r c_j c_k}\right),\label{eq:5}
\end{align}
respectively.

 \section{Results}

\begin{figure}[t]
\begin{center}
\includegraphics[width=\columnwidth,clip]{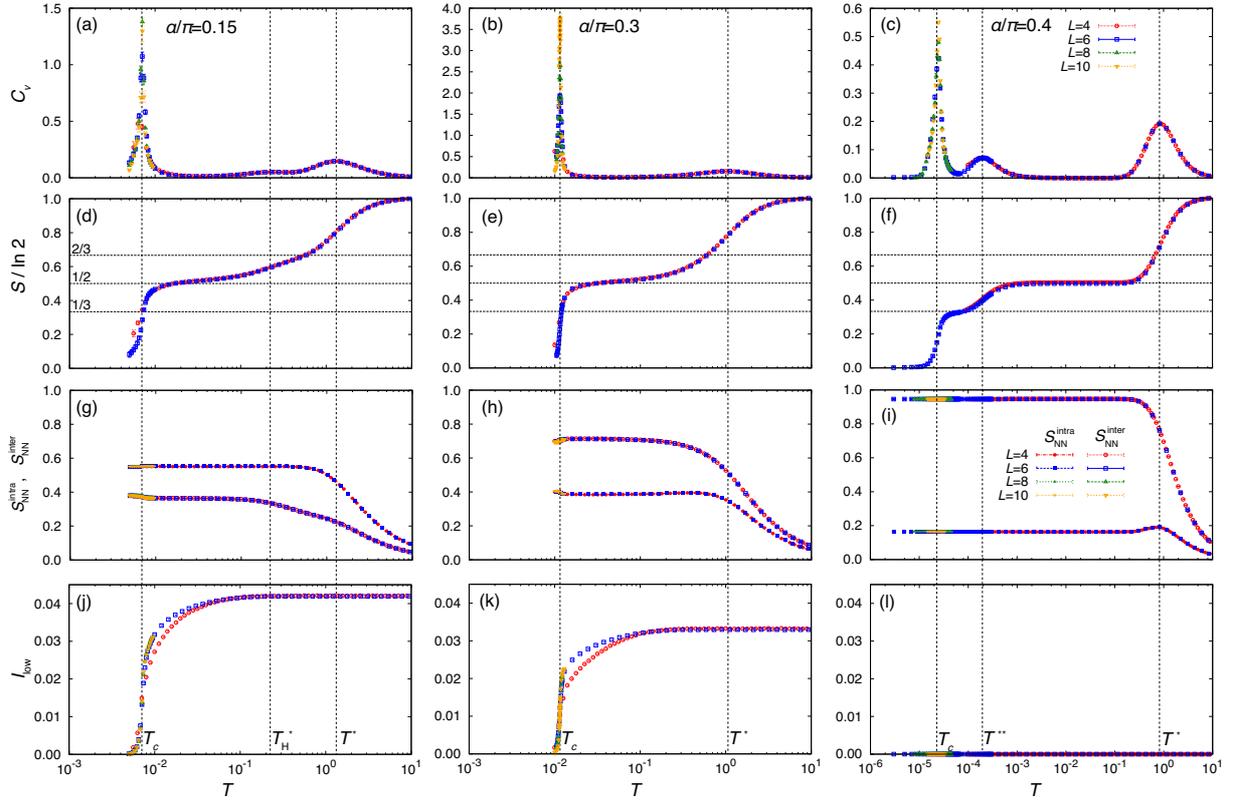}
\caption{
 (a)--(c) Temperature dependences of the specific heat at (a) $\alpha/\pi=0.15$, (b) $\alpha/\pi=0.3$, and (c) $\alpha/\pi=0.4$ for several cluster sizes [see Eq.~(\ref{eq:3})].
 (d)--(f) The corresponding data of the entropy per site normalized by $\ln 2$ [see Eq.~(\ref{eq:7})].
 (g)--(i) The NN spin correlations $S_{\rm NN}^{\rm intra}$ on the intra-triangle $\gamma$ bonds with the exchange constant $J$ and $S_{\rm NN}^{\rm inter}$ on the inter-triangle $\gamma'$ bonds with the exchange constant $J'$ [see Eqs.~(\ref{eq:4}) and (\ref{eq:5})].
 (j)--(l) The low-energy weight of the DOS between 0 and $\Delta_0/2$, $I_{\rm low}$, where $\Delta_0$ is the excitation gap at $T=0$ [see Eq.~(\ref{eq:I_low})].
 The vertical lines represent the transition temperature $T_c$ and the crossover temperatures $T_{\rm H}^*$, $T^*$, and $T^{**}$.
 }
\label{corr}
\end{center}
\end{figure}

Figure~\ref{corr}(a) shows the specific heat for several size clusters at $\alpha/\pi=0.15$, where the system is in the topologically-nontrivial CSL region.
There is a sharp peak growing with the system size at $T_c\simeq 0.007$.
This indicates that the phase transition occurs at $T_c$, which is associated with the time reversal symmetry breaking: the CSL phase is stabilized below $T_c$~\cite{PhysRevLett.115.087203}.
Through this phase transition, a half of the entropy $S$ is released as shown in Fig.~\ref{corr}(d).
The entropy release comes from the coherent development of both $W_{\rm t}$ and $W_{\rm d}$~\cite{PhysRevLett.115.087203}: the sum of the numbers of $W_{\rm t}$ and $W_{\rm d}$ is a half of the number of spins.
Above $T_c$, however, the specific heat shows two broad peak structures at 
$T_{\rm H}^*\simeq 0.25$ and $T^*\simeq 1.2$.
These are almost independent of the system size, and hence, correspond to crossovers~\cite{PhysRevLett.115.087203}. 
(We follow the notations for the crossover temperatures in Ref.~\cite{PhysRevLett.115.087203}.)
As shown in Fig.~\ref{corr}(d), the entropy is released by $\frac13 \ln 2$ and $\frac23 \ln 2$ around $T_{\rm H}^*$ and $T^*$,
respectively.

In order to reveal the origin of the entropy releases, we calculate the spin correlations for the NN bonds, $S_{\rm NN}^{\rm intra}$ and $S_{\rm NN}^{\rm inter}$.
Figure~\ref{corr}(g) shows $S_{\rm NN}^{\rm intra}$ and $S_{\rm NN}^{\rm inter}$ as functions of $T$ at $\alpha/\pi=0.15$.
$S_{\rm NN}^{\rm intra}$ develops rapidly around $T^*$, and saturates to the $T=0$ value with decreasing $T$.
On the other hand, $S_{\rm NN}^{\rm inter}$ shows successive growth around $T_{\rm H}^*$ and $T^*$ before the saturation.
These contrastive $T$ dependences are understood by the difference of the exchange constants $J$ and $J'$.
As $J$ is larger than $J'$ at $\alpha/\pi=0.15$, the intra-triangle correlation $S_{\rm NN}^{\rm intra}$ develops at a higher $T \sim T^*$, while the inter-triangle one $S_{\rm NN}^{\rm inter}$ grows at a lower $T \sim T_{\rm H}^*$.
 The growth of $S_{\rm NN}^{\rm inter}$ around $T^*$ is induced by the development of $S_{\rm NN}^{\rm intra}$.
Although the spin correlations show large development through the crossovers, they exhibit a slight change in the vicinity of $T_c$, where thermal fluctuations of the $Z_2$ variables $\{\eta_r \}$ are enhanced.
This result indicates that the NN spin correlations are not affected by the $Z_2$ fluctuations significantly.
In other words, as the NN spin correlations are nothing but the kinetic energy of the itinerant Majorana fermions as shown in Eqs.~(\ref{eq:4}) and~(\ref{eq:5}), the kinetic motion of the Majorana fermions is not disturbed by the fluctuations of the localized Majorana fermions.
This behavior is common in other parameter region, as shown in Figs.~\ref{corr}(h) and~\ref{corr}(i).

Next, we focus on the case of $\alpha/\pi=0.3$, which is still in the topologically-nontrivial CSL region but closer to the critical point at $\alpha_c/\pi=1/3$.
Figure~\ref{corr}(b) shows the $T$ dependence of the specific heat.
The result shows a sharp peak at $T_c\simeq 0.01$ and a broad peak at $T^*\simeq 1$.
The $T$ dependence of the entropy in Fig.~\ref{corr}(e) indicates that the entropy is released by halves successively at $T_c$ and $T^*$.
This result implies that the two crossovers $T_{\rm H}^*$ and $T^*$ at $\alpha/\pi=0.15$ merges into a single crossover at $T^*$ with increasing $\alpha$.
This is also consistent with the behavior of the NN spin correlations:
$S_{\rm NN}^{\rm intra}$ and $S_{\rm NN}^{\rm inter}$ grow together around $T^*$, as shown in Fig.~\ref{corr}(h). 
The difference between the $T$ dependences of the NN spin correlations at $\alpha/\pi=0.15$ and $0.3$ is understood from the DOS of the Majorana fermions, as discussed later.

With further increasing $\alpha$, the ground state changes from the topologically-nontrivial CSL to the topologically-trivial CSL at $\alpha_{c}/\pi=1/3$.
Figure~\ref{corr}(c) shows the specific heat in the topologically-trivial CSL region at $\alpha/\pi=0.4$.
Similar to the result at $\alpha/\pi=0.15$ in Fig.~\ref{corr}(a), there appear again two broad peaks associated with crossovers, in addition to a sharp peak corresponding to the phase transition.
However, at $\alpha/\pi=0.4$, a half of the entropy is released around the high-$T$ crossover at $T\sim 0.8$, as shown in Fig.~\ref{corr}(f).
Therefore, this crossover corresponds to that at $T^*$ for $\alpha/\pi=0.3$.
Meanwhile, the low-$T$ crossover at $T^{**}\sim 0.0002$ does not correspond to that at $T_{\rm H}^*$ for $\alpha/\pi=0.15$.
This crossover is associated with the coherent growth of the conserved quantity $W_{\rm d}$~\cite{PhysRevLett.115.087203,Nasu2015pre}.
Figure~\ref{corr}(i) shows the $T$ dependences of the NN spin correlations $S_{\rm NN}^{\rm intra}$ and $S_{\rm NN}^{\rm inter}$.
Both of them grow with decreasing $T$ around $T^*$, while $S_{\rm NN}^{\rm intra}$ shows a hump structure.
However, they show no significant change at $T^{**}$ as well as $T_c$.

\begin{figure}[t]
\begin{center}
\includegraphics[width=0.7\columnwidth,clip]{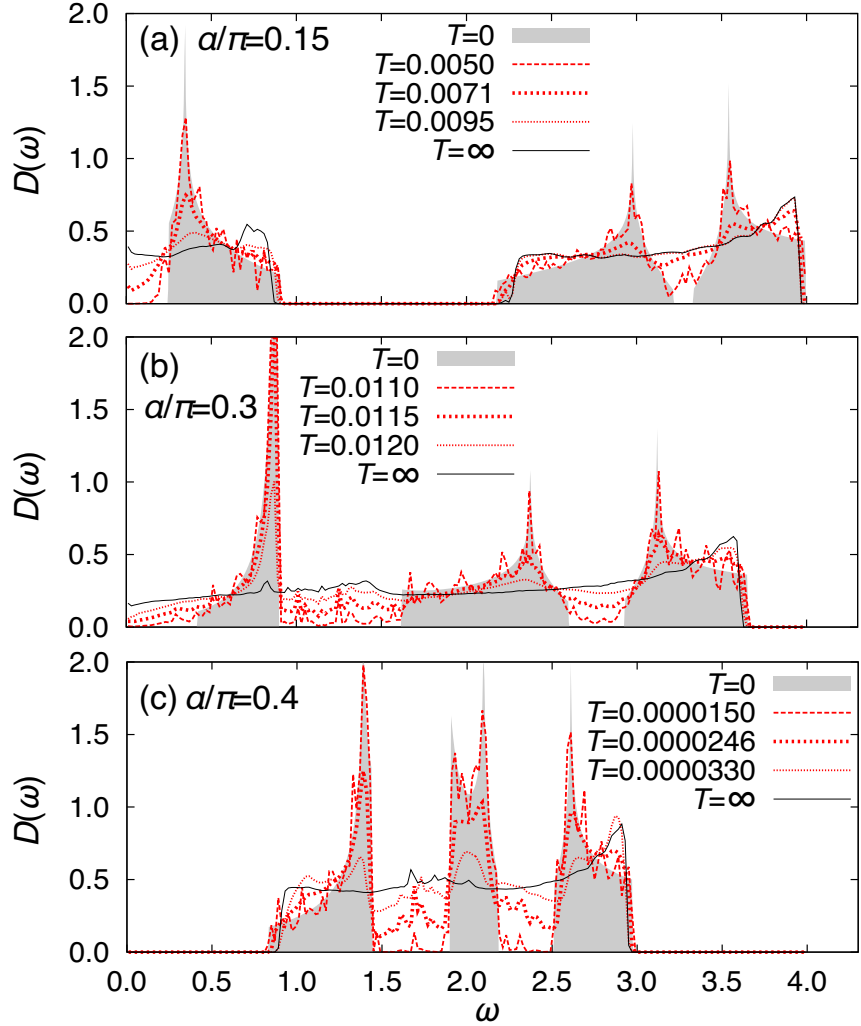}
 \caption{
 The DOS of the itinerant Majorana fermions, $D(\omega)$, at (a) $\alpha/\pi=0.15$, (b) $\alpha/\pi=0.3$, and (c) $\alpha/\pi=0.4$ for several temperatures around $T_c$, which are obtained by QMC for $L=10$ [see Eq.~(\ref{eq:DOS})].
 For comparison, the results at $T=0$ and $T=\infty$ are also plotted.
 The DOS at $T=\infty$ are obtained from 1000 random configurations of $\{\eta_r\}$ in the $L=20$ cluster.
 }
\label{dos}
\end{center}
\end{figure}

As mentioned above, the spin correlation is given by the kinetic energy of the itinerant Majorana fermions in the current system.
To gain further insight into the magnetic properties, we calculate the $T$ evolution of the DOS of the itinerant Majorana fermions.
Figure~\ref{dos} shows the $T$ dependence of the DOS $D(\omega)$ at $\alpha/\pi=0.15$, $0.3$, and $0.4$.
At $T=0$, the spectrum is split into three parts for all the cases, corresponding to the three bands of the itinerant Majorana fermions.
In the case of $\alpha/\pi=0.15$, the lowest-energy band is largely separated from the two higher-energy bands at $T=0$, as shown in Fig.~\ref{dos}(a).
With increasing $T$, the gap between the two higher-energy bands is filled around $T_c$, and they merge into a single band above $T_c$.
On the other hand, the gap between the lowest-energy band and the higher-energy one remains robust even at $T=\infty$.
In contrast, at $\alpha/\pi=0.3$ and $0.4$, all the three Majorana fermion bands merge into a single one above $T_c$ with increasing $T$, as shown in Figs.~\ref{dos}(b) and ~\ref{dos}(c).

These overall evolutions of the DOS play a crucial role in the thermodynamics in the paramagnetic state above $T_c$.
In the case of $\alpha/\pi=0.15$, there appear two bands above $T_c$.
This explains why the system exhibits two crossovers above $T_c$ as follows.
Suppose the DOS is approximated by a constant where it is nonzero in Fig.~\ref{dos}(a) for $T > T_c$, namely, $D(\omega) = D_{\rm low}$ for $0<\omega\lesssim0.8$ and $D(\omega) = D_{\rm high}$ for $2.3\lesssim\omega\lesssim 4$, the contributions from $D_{\rm low}$ and $D_{\rm high}$ to the specific heat through Eq.~(\ref{eq:6}) shows a maximum at $T\sim 0.2$ and $T\sim 1.3$, respectively.
These temperatures agree well with $T_{\rm H}^*$ and $T^*$, respectively.
Thus, the two well-separated bands above $T_c$ explain the two crossovers, where the NN spin correlations $S_{\rm NN}^{\rm inter}$ and  $S_{\rm NN}^{\rm intra}$ are saturated.
On the other hand, at $\alpha/\pi=0.3$ and $0.4$, there is only a single band above $T_c$. 
This leads to a single crossover at $T^*$, where $S_{\rm NN}^{\rm inter}$ and  $S_{\rm NN}^{\rm intra}$ develop simultaneously with decreasing $T$.

Next, we discuss the low-energy behavior in the DOS.
In the case of $\alpha/\pi=0.15$ and $0.3$, the energy gap to the lowest excited state is filled with increasing $T$ in the vicinity of $T_c$, as shown in Figs.~\ref{dos}(a) and \ref{dos}(b), respectively.
On the other hand, as shown in Fig.~\ref{dos}(c), the low-energy excitation gap remains open at $\alpha/\pi=0.4$.
In order to see the $T$ dependence of the gap more clearly, we calculate the low-energy weight of the DOS, $I_{\rm low}$, which is defined by
\begin{align}
\label{eq:I_low}
 I_{\rm low}=\int_0^{\Delta_0/2}D(\omega)d\omega,
\end{align}
where $\Delta_0$ represents the excitation gap at $T=0$.
Figure~\ref{corr}(j) shows the $T$ dependence of $I_{\rm low}$ at $\alpha/\pi=0.15$.
The low-energy weight abruptly increases at $T_c$ with increasing $T$.
This behavior is also seen in the case of $\alpha/\pi=0.3$ as shown in Fig.~\ref{corr}(k).
However, as presented in Fig.~\ref{corr}(l), $I_{\rm low}$ always vanishes even at high $T$ in the case of $\alpha/\pi=0.4$.
These results indicate that the excitation gap is filled by introducing thermal fluctuations of the $Z_2$ variables in the topologically-nontrivial CSL region.
Namely, the low-energy gap in the non-Abelian CSL phase is fragile against thermal fluctuations of the $Z_2$ variables, whereas the gap in the Abelian CSL phase is robust.
The gap closing in the former takes place at $T_c$.
Therefore, in the non-Abelian CSL phase below $T_c$, the topological number of the Majorana fermion bands is well-defined and topological transports such as the thermal Hall effect will be observed~\cite{PhysRevLett.115.087203}.
Note that, in the present calculations, it is difficult to distinguish whether the gap closing is due to the thermal fluctuation of $W_{\rm t}$ or that of $W_{\rm d}$ because the associated entropies are released simultaneously at $T_c$ in the cases of $\alpha/\pi=0.15$ and $0.3$.
It is an intriguing issue to clarify which conserved quantity is relevant to the gap closing in the non-Abelian CSL.
This is left for a future work.

\begin{figure}[t]
\begin{center}
\includegraphics[width=0.7\columnwidth,clip]{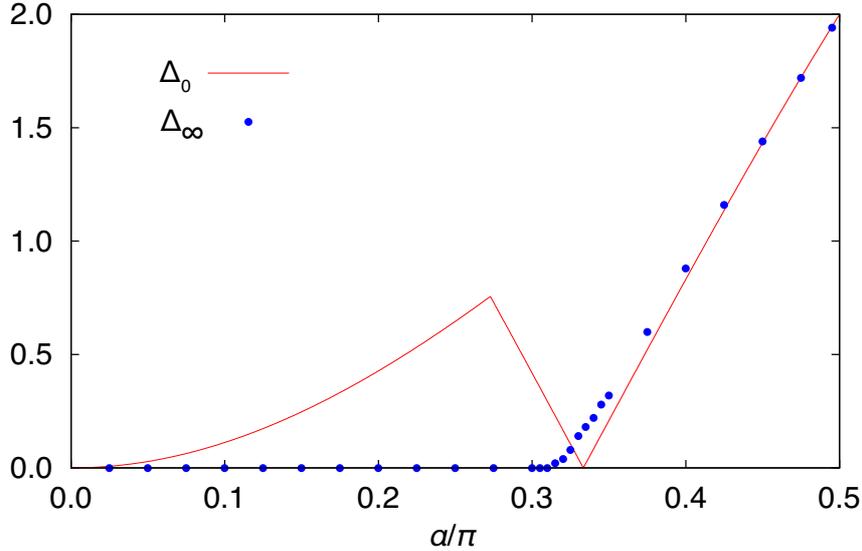}
\caption{
 The excitation gaps in the itinerant Majorana fermion spectrum as functions of $\alpha$: $\Delta_0$ and $\Delta_\infty$ are the values at $T=0$ and $T=\infty$, respectively.
 }
\label{gap}
\end{center}
\end{figure}

Finally, in order to clarify the $\alpha$ dependence of the excitation gap in the paramagnetic phase above $T_c$, we calculate the gap in the high-$T$ limit.
This is obtained from 1000 random configuration of $\{\eta_r\}$ in the $L=20$ cluster.
Figure~\ref{gap} shows the gap estimated from the DOS at $T=\infty$, $\Delta_\infty$, as a function of $\alpha$, together with the value at $T=0$, $\Delta_0$~\cite{PhysRevLett.99.247203}.
As mentioned before, $\Delta_0$ vanishes at $\alpha_c/\pi=1/3$ associated with the change of the topological nature in the Majorana fermion band.
At $T=\infty$, the excitation gap disappears in the non-Abelian CSL region for $\alpha \lesssim \alpha_c$, but it does not change largely in the Abelian CSL region for $\alpha \gtrsim \alpha_c$.
(We note that the gapless-gapped boundary at $T=\infty$ is slightly smaller than $\alpha_c$.)
This contrastive behavior is one of striking characteristics reflecting the topology of the Majorana fermion band even in the paramagnetic state above $T_c$.

\section{Summary}
 
  To summarize, we have investigated the finite-$T$ properties in the Kitaev model on the decorated honeycomb lattice, with focusing on the magnetic properties and the dynamics for the Majorana fermion system.
  We found that the spin correlations increase with decreasing $T$ around the crossovers in the paramagnetic state associated with the entropy release of the itinerant Majorana fermions.
  The NN correlations in the intra- and inter-triangle bonds develop at different $T$ in the small $\alpha$ region, whereas they develop simultaneously when $\alpha$ becomes larger.
  We also found that the DOS of the itinerant Majorana fermions changes largely at $T_c$, and the spectrum for $T>T_c$ gives an explanation for the crossovers with the changes of magnetic properties.
Furthermore, we revealed that the excitation gap vanishes above $T_c$ in the non-Abelian CSL region, while it remains open in the Abelian CSL region.

\ack
This work is supported by Grant-in-Aid for Scientific Research No.~15K13533, 
the Strategic Programs for Innovative Research (SPIRE), MEXT, and the Computational Materials Science Initiative (CMSI), Japan.
Parts of the numerical calculations are performed in the supercomputing systems in ISSP, the University of Tokyo.

\section*{References}

\bibliographystyle{iopart-num}
\bibliography{refs}

% \end{thebibliography}
% \smallskip

\end{document}